\documentclass[aps,prb,amsmath,amssymb,twocolumn,groupedaddress]{revtex4}

\usepackage{graphicx}% Include figure files
\usepackage{dcolumn}% Align table columns on decimal point
\usepackage{bm}% bold math

\begin{document}

%\preprint{}

\title{Lattice Vibrational Modes in Si/Ge Core-shell Nanowires}


\author{Shouting Huang and Li Yang}
%\thanks{}
\affiliation{Department of Physics, Washington University, St.
Louis, Missouri 63130, USA}


\date{\today}

\begin{abstract}
We present a first-principles study on lattice vibrational modes
of Si/Ge core-shell nanowires (NWs). In addition to quantum
confinement, the internal strain induced by the lattice mismatch
between core and shell contributes to significant frequency shifts
of characteristic optical modes. More importantly, our simulation
shows that these frequency shifts can be detected by Raman
scattering experiments, providing convenient and nondestructive
ways to obtain structural information of core-shell materials.
Meanwhile, another type of collective modes, radial breathing
modes (RBMs), are identified in Si-core/Ge-shell NWs and their
frequency dependence is explained by an elastic media model.
\end{abstract}

%\keywords{}

\maketitle

\section{\label{sec1}INTRODUCTION}

Core-shell nanowires (NWs), a type of novel quasi-one-dimensional
nanostructures, have ignited tremendous interest to date because
these unique radial heterojunctions own extra degrees of freedom
to tune the mechanical, electric and optical properties by varying
their core and shell sizes, respectively, which make them superior
to usual homogenous materials
\cite{2002Lauhon,2006Xiang,2007Tian,2008Zhang}. Among numerous
core-shell structures, Si/Ge core-shell NWs are of particular
interest because of the known importance of silicon on
microelectronics. Recent research has shown that Si/Ge core-shell
NWs own a number of unusual thermal transport features for energy
applications \cite{2011HuNL,2011Chen,2011HuPRB}. For example, the
enhanced difference in thermal and electric transport behaviors
between the core and shell gives rise to higher-efficiency
thermoelectric devices based on core-shell NWs \cite{2011Wingert}.

In order to realize and optimize the above interesting properties,
it is imperative to determine how to obtain structural information
and, thereafter, how to design core and shell geometries.
Traditional approaches, such as the transmission electron
microscopy (TEM) and neutron scattering measurements, are time
consuming or expensive for the large-scale production; alternative
convenient ways are highly desired. On the other hand, lattice
vibrational modes are known to be sensitive to the chemical
bonding and boundary conditions of nanostructures and they are
well accessible by varieties of experiments, such as Raman
scattering \cite{2000Wang,2005Adu,2011Yang}, making it possible to
study lattice vibrational modes of core-shell NWs and extract
useful structural information.

Beyond application interests, lattice vibrational modes and
corresponding phonons are fundamental properties of solids.
However, compared with bulk semiconductors and homogenous
nanostructures, there have been very limited first-principles
studies on vibrational modes of core-shell NWs although the
core-shell structure will surely introduce richer physics.
Therefore, calculating lattice vibrational modes in Si/Ge
core-shell NWs will be an appropriate starting point to understand
electric and thermal properties of reduced dimensional structures.

In this work, we employ first-principles calculations and focus on
the following two topics of Si/Ge core-shell NWs: 1) how the
unique core-shell structure modifies lattice vibrational modes and
what their impacts on thermal properties are; 2) how to extract
the structural information from characters of lattice vibrational
modes and provide clues for experimental measurements.

Through our simulations, we find that the structural information
of core-shell NWs can be efficiently obtained by frequency shifts
of high-energy optical vibrational modes. This type of frequency
shifts can be well explained by the change of the strain induced
by the lattice mismatch between the core and shell. Moreover, we
have calculated the Raman scattering spectrum of Si/Ge core-shell
NWs, showing that these frequency shifts can be detected by
experiments. Meanwhile, we identify the existence of RBMs in
Si-core/Ge-shell NWs, whose frequency variation can be understood
by an elastic media model.

%Our analysis shows that the core-shell structure can enhance the
%difference of thermal properties of core and shell, providing
%promising evidence to improve thermoelectric efficiency of
%core-shell nanostructures.

The rest of this article is organized as the following. In
Sec. \ref{sec2}, we present our calculation methodology and
computational setup. In Sec. \ref{sec3}, we discuss the simulation
results, including properties of high-frequency optical modes,
Raman scattering spectra, and the RBM frequency dependence on the
structure of core-shell NWs. In Sec. \ref{sec4}, we summarize our
studies and conclusion.

\section{\label{sec2}COMPUTATIONAL DETAILS}

Our calculation is based on the density functional theory (DFT)
within the local density approximation (LDA)
\cite{1964Hohenberg,1965Kohn} as implemented in the Quantum
ESPRESSO package \cite{2009Giannozzi}. We employ the plan-wave
basis and the pseudopotential approximation \cite{1991Troullier}.
The plane-wave energy cutoff is 16 Ry. For the Brillouin zone (BZ)
integration, we use a 1$\times$1$\times$8 k-point sampling grid.
All calculations are done in a supercell arrangement
\cite{1975Cohen} with a 0.7-nm vacuum distance between neighboring
NWs. Our studied NWs are oriented along the [110] direction, an
energetically preferred configuration \cite{2004Wu}. Those
dangling bonds on the surface of NWs are passivated by hydrogen
atoms. All NWs are fully relaxed according to the force and stress
calculated under DFT/LDA.

\begin{table*}
\caption{\label{table1}The structure information of Si/Ge
core-shell NWs and homogenous NWs investigated.(SiGe means Si core
and Ge shell while GeSi means Ge core and Si shell.)}
\begin{ruledtabular}
\begin{tabular}{cccccc}
NWs&Core&Shell&Core diameter (nm)
&Shell diameter (nm)&Lattice constant along wire (nm)\\
\hline
 Si-1&Si&Si&2.24&2.24&0.378 \\
 GeSi-1&Ge&Si&1.12&2.21&0.383 \\
 GeSi-2&Ge&Si&1.75&2.30&0.390 \\
 Ge-1&Ge&Ge&2.41&2.41&0.401 \\
 SiGe-1&Si&Ge&1.09&2.33&0.396 \\
 SiGe-2&Si&Ge&1.66&2.29&0.388 \\
 Si-2&Si&Si&1.61&1.61&0.382 \\
 Ge-2&Ge&Ge&1.76&1.76&0.405 \\
\end{tabular}
\end{ruledtabular}
\end{table*}

There are a number of degrees of freedom to decide the structure
of core-shell NWs, such as the core size, shell size and the
chemical compositions. Because it is impossible to study all of
them in one article, we investigate Si/Ge core-shell NWs with a
fixed diameter ($\sim$ 2.3 nm) but a varying core-shell size
ratio. At the same time, homogenous NWs are also studied for
comparison purposes. The detailed structural information of our
studied NWs is reported in Table \ref{table1}.

We employ the linear response approach to obtain lattice
vibrational modes and their frequencies
\cite{1986Baroni,1987Baroni,1991Giannozi,1992Gonze}. Only
vibrational modes at the $\Gamma$ point of the first BZ is
calculated because it is of particular interest for the
first-order Raman scattering measurement.

We calculate the Raman scattering spectrum by considering the
non-resonant first-order process, which can be described within
the Placzek approximation \cite{1982Cardona,1986Bruesch}. Then the
Raman activity $I_{ramman}^{k}$ associated with the vibrational
mode $k$ is

\begin{equation}\label{equ1}
I_{ramman}^{k} = |e_{s}\cdot \alpha^{k} \cdot e_{L}|,
\end{equation}
where $e_{s}$ is the polarization of incoming photon and $e_{L}$
is the polarization of scattered outgoing photon. The Raman tensor
$\alpha^{k}$ is defined as

\begin{equation}\label{equ2}
\alpha^{k} = \sqrt[2]{\Omega} \sum_{al} \frac{\partial
\chi_{ij}}{\partial r_{al}} u_{al}^{k},
\end{equation}
where $\chi_{ij}$ is the electric polarizability tensor, $a$ is
the index to specify the atom in the unit cell, and $l$ is the
index of coordinates. $u_{al}^{k}$ is the displacement of the atom
$a$ along $l$ direction in the vibrational eigenmode $k$.

For realistic experimental cases, NWs are usually oriented
randomly. Because of the depolarization effect, we may only
consider the circumstances that the direction of the incident
beam, the polarization direction of this beam, and the direction
of the observation are perpendicular to each other. Then the
spatially averaged Raman activity of those randomly orientated NWs
is given by:

\begin{widetext}
\begin{equation}\label{equ3}
I_{ramman}^{k} = 5(\alpha_{xx}^{k} + \alpha_{yy}^{k} +
\alpha_{zz}^{k})^{2} + \frac{7}{4}[(\alpha_{xx}^{k} -
\alpha_{yy}^{k})^{2} + (\alpha_{xx}^{k} - \alpha_{zz}^{k})^{2} +
(\alpha_{yy}^{k} - \alpha_{zz}^{k})^{2}
 + 6(|\alpha_{xy}^{k}|^{2} + |\alpha_{xz}^{k}|^{2} +
 |\alpha_{yz}^{k}|^{2})].
\end{equation}
\end{widetext}

Because our calculated Raman spectra only consider the first-order
process, they may be different from some Raman experiments which
could be dominated by resonant effects. However, our obtained
Raman activity, \emph{e.g.}, active or not, shall still be useful
to understand those resonant experiments.

\section{\label{sec3}RESULTS AND DISCUSSIONS}

\subsection{High-frequency optical modes}

\begin{figure}
\centering
\includegraphics[width=1.0\columnwidth]{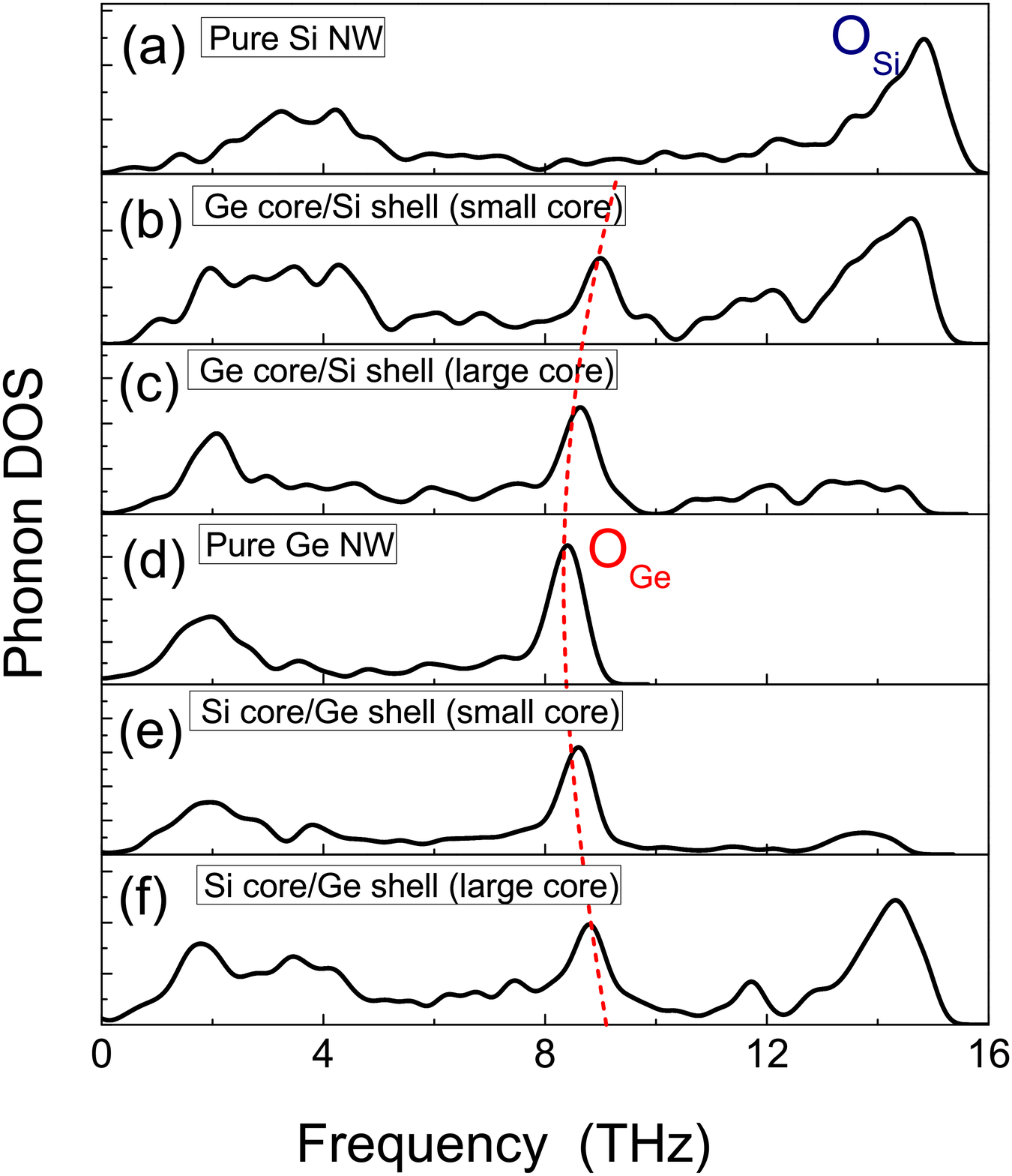}%
\caption{\label{fig1} DVM at the $\Gamma$ point of Si/Ge
core-shell NWs and homogenous NWs with wire diameter of about 2.3
nm in Table \ref{table1}. (a) Si-1 NW (b) GeSi-1 NW (c) GeSi-2 NW
(d) Ge-1 NW (e) SiGe-1 NW (f) SiGe-2 NW. The red dashed line is
used to guide the shift trend of the peak of high-frequency Ge-Ge
optical modes.}
\end{figure}

The density of vibrational modes (DVM) at the $\Gamma$ point is
presented in Fig. \ref{fig1}. It has to be pointed out that those
vibrational modes mainly consisting of hydrogen-atom motions are
excluded from our plots because we are interested in intrinsic
vibrational modes of NWs that are not sensitive to the
environment. Therefore, our results and conclusion will be more
universal and less influenced by complicated passivations as in
various experiments.

\begin{figure}
\centering
\includegraphics[width=1.0\columnwidth]{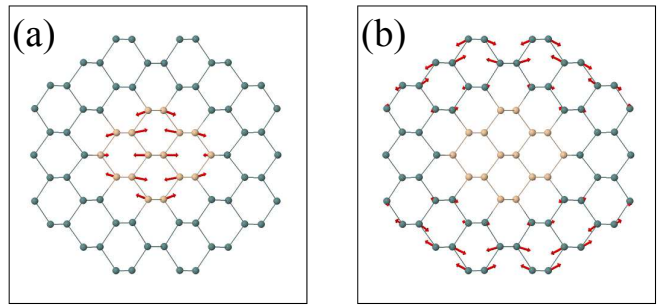}%
\caption{\label{fig2} The typical vibrations of highest-frequency
optical modes in Si/Ge core-shell NWs. Hydrogen atoms on the
surface are not plotted. The arrows stand for the motion direction
and magnitude. (a) The highest-frequency Si optical mode in the
$O_{Si}$ peak of a Si-core/Ge-shell NW. (b) The highest-frequency
Ge optical mode in the $O_{Ge}$ peak of the Si-core/Ge-shell NW.}
\end{figure}

The most prominent feature of these DVMs is that there are two
characteristic peaks among most core-shell NWs, which are marked
by $O_{Si}$ and $O_{Ge}$ in Fig. \ref{fig1}, respectively. To
understand the physical origin of these peaks, we have plotted the
real-space motion of two typical modes corresponding to these
peaks in Fig. \ref{fig2}.

The first character of these modes is that they are mainly
confined within the core or shell, respectively. In another word,
these Si and Ge vibrational modes are decoupled. For example, Fig.
\ref{fig2} (a) is mainly the Si-Si vibrational mode confined
within the core regime while Fig. \ref{fig2} (b) is mainly the
Ge-Ge mode within the shell regime. This decoupling is due to
the significant mass difference between Si and Ge atoms and
their bonding strength differences.

The second character is that the neighboring atoms of these modes
plotted in Fig. \ref{fig2} have opposite vibrational directions, a
typical feature of optical modes of semiconductors at the $\Gamma$
point. Based on the above two characters, we conclude that these
peaks ($O_{Si}$ and $O_{Ge}$) originate from those high-frequency
optical modes of bulk Si and Ge. This can be further confirmed by
their frequency; the $O_{Ge}$ peak has a frequency around 8 THz
while the $O_{Si}$ peak has a frequency around 15 THz. They are
consistent with frequencies of optical modes in Si and Ge bulk
crystals.

However, the frequencies of peaks ($O_{Ge}$ and $O_{Si}$) are not
exactly the same as their bulk counterparts. Instead they exhibit
substantial frequency shifts as the core-shell size varies as
shown in Fig. \ref{fig1}. For example, as we expand the Ge-core
size from zero to the whole Ge-core/Si-shell wire (Figs.
\ref{fig1} (a) to (d)), both peaks $O_{Si}$ and $O_{Ge}$ have a
red shift. Similarly, both peaks have a blue shift as Si core size
increases in the case of Si-core/Ge-shell NWs, as shown in Figs.
\ref{fig1} (d) to (f).

The physical origin of the above shifts of prominent peaks in the
DVM can be attributed to two factors, quantum confinement and
strain condition. First, quantum confinement in such narrow NWs
definitely modifies the frequency of lattice vibrational modes as
discovered by previous studies
\cite{2000Wang,2005Adu,2009peeters,2011Yang}. However, since the
whole diameter of our studied NWs is fixed, quantum confinement
shall not be a major factor for the frequency shift observed in
Fig. \ref{fig1}.

On the other hand, the internal strain of core-shell NWs induced
by lattice mismatch can be an important reason for those frequency
shifts. In our Si/Ge core-shell NWs, the Si part is stretched
while the Ge part is compressed. As the core-shell size changes,
the average lattice constant along the axial direction varies
according to the Vegard's law ($a=a_{A}^{0}(1-X)+a_{B}^{0}X$,
where $a$ is the lattice parameter of $A_{1-X}B_{X}$ crystal, X is
the concentration of constituent element B, $a_{A}^{0}$ and
$a_{B}^{0}$ are the lattice parameters of pure A and pure B
crystals, respectively.), which can be seen from Table I. As a
result, the intrinsic strain will gradually change, resulting
in significant frequency shifts of Si and Ge modes.

\begin{figure}
\centering
\includegraphics[width=1.0\columnwidth]{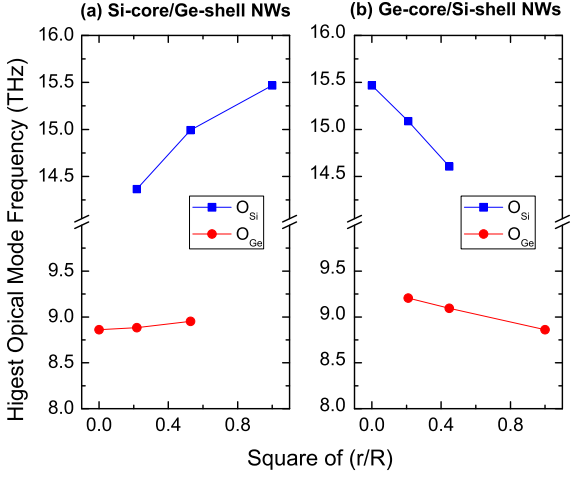}%
\caption{\label{fig3} The highest frequency of the optical mode
picked from $O_{Si}$ and $O_{Ge}$ peaks. $r$ is the core radius
and $R$ is the whole wire radius. (a) Si-core/Ge-shell NWs with a
fixed wire diameter. (b) Ge-core/Si-shell NWs with a fixed wire
diameter.}
\end{figure}

To further confirm the above explanation, we focus on the
frequency shifts of the particular optical modes picked from
$O_{Si}$ and $O_{Ge}$ peaks, respectively. As shown in Fig.
\ref{fig3}, for both Si-core/Ge-shell and Ge-core/Si-shell NWs,
these frequency shifts are approximately linear to the square of
the ratio of the core and NW radii, which is also proportional to
the ratio of the number of atoms in the core and the whole NW.
Since this linear relation is exactly the lattice constant
variation according to the Vegard's Law, the change of the
frequency shall be a result of the variation of the internal
strain.

\begin{figure}
\centering
\includegraphics[width=1.0\columnwidth]{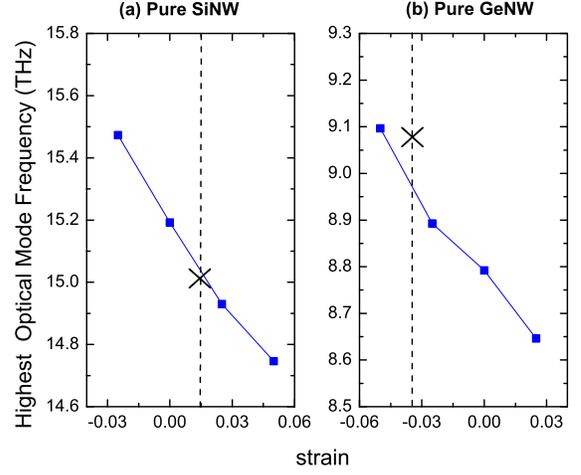}%
\caption{\label{fig4} Highest optical mode frequencies with
external uniaxial strains applied to pure Si NW and pure Ge NW
respectively. (a) Si-2 NW, the cross marks the frequency position
of the highest Si optical mode in SiGe-2 NW. (b) Ge-2 NW, the
cross marks the frequency position of the highest Ge optical mode
in GeSi-2 NW.}
\end{figure}

Meanwhile, we calculate homogenous Si NWs (SiNWs) and Ge NWs (GeNWs)
with applied axial strain to check the strain effect as shown in Fig.
\ref{fig4}. The size of homogenous NWs is chosen to be similar to
the corresponding core size of Si/Ge core-shell NWs (SiGe-2 and
GeSi-2 NWs in Table I) and we focus on the highest-frequency
optical mode, which is the same mode plotted in Fig. \ref{fig3}.
For both SiNW and GeNWs, these modes exhibit a monotonic change
with the applied axial strain; their frequencies increase with the
compressive strain and decrease with the tensile strain, which is
consistent with the trend of the peak shifts in core-shell NWs.
Moreover, we mark the highest-frequency optical mode of the
corresponding core-shell NWs in Fig. \ref{fig4}. We find that
their frequencies are close to those of the highest-frequency optical mode
of homogenous NWs under the similar strain. Therefore, the
variation of the strain condition is the main reason for the
frequency shifts of high-frequency optical modes in core-shell
NWs. This result builds a bridge to connect the frequency shift
of characteristic optical modes with the structure of core-shell
NWs.

%\subsection{Transverse and longitudinal optical modes splitting}
%
%Strain effect not only causes frequency shifts of optical modes
%but also affects the longitudinal and transverse optical modes
%splitting (TO-LO splitting). This TO-LO splitting is observed in
%Si/Ge core-shell NWs, as shown in Fig. \ref{figsplit}. Here we
%only plot the highest-frequency TO, which is always vibrating
%along the [001] direction. In Fig. \ref{figsplit}, the splitting
%varies as the core shell structure changes. This change of
%splitting is a result of competition of internal strain and
%quantum confinement.
%
%In previous studies \cite{2011Yang}, quantum confinement effect
%will enhance the TO-LO splitting. As a result, we observe this
%splitting for homogenous NWs in Fig. \ref{figsplit}, where
%$r/R=1$. On the other hand, for example, in Fig. \ref{figsplit}
%(a), as the size of the Si core increases, the strain has
%different impacts on TO and LO. As a result, the TO-LO splitting
%is no longer a simple monotonic relation to the core-shell
%geometry.
%
%\begin{figure}
%\centering
%\includegraphics[width=1.0\columnwidth]{splitting.eps}%
%\caption{\label{figsplit} Highest TO-LO mode splitting in core-shell NWs.
%(a) core modes splitting in Si-core/Ge-shell NWs. (b) core modes splitting
%in Ge-core/Si-shell NWs.}
%\end{figure}

\subsection{Raman scattering spectrum}

\begin{figure}
\centering
\includegraphics[width=1.0\columnwidth]{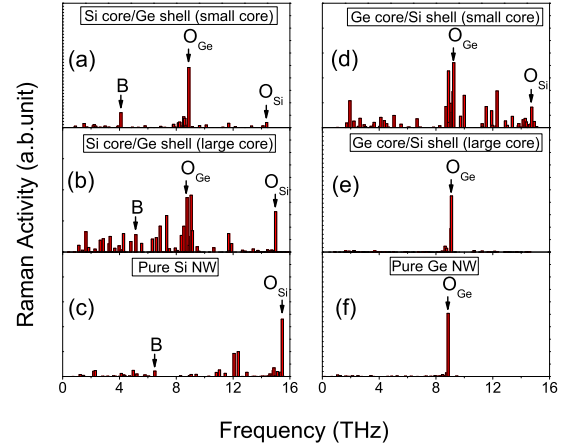}%
\caption{\label{fig5} Raman activity of Si/Ge core-shell NWs and
homogenous NWs. (a) SiGe-1 NW. (b) SiGe-2 NW. (c) Si-1 NW. (d)
GeSi-1 NW. (e) GeSi-2 NW. (f) Ge-1 NW. $O_{Si}$ marks the
highest-frequency Si optical mode, $O_{Ge}$ marks the
highest-frequency Ge optical mode, and $B$ marks the RBM.}
\end{figure}

Raman scattering is a widely used approach to detect lattice
vibrational modes of solids. Recently it is calculated in SiNWs as
well \cite{2011Yang, 2005Mahan}. If the above frequency shifts of
characteristic optical modes can be observed in their Raman
scattering spectrum, it will be useful to understand the
structures of core-shell NWs. Following the formulas in
Eq.~(\ref{equ2}) and Eq.~(\ref{equ3}), we calculate the
first-order Raman scattering spectra of our studied Si/Ge
core-shell NWs and present them in Fig. \ref{fig5}.

In Fig. \ref{fig5}, we can see the following important features.
First, those highest-frequency optical modes in the peaks $O_{Si}$
and $O_{Ge}$ are strongly active in the Raman scattering spectra.
This is reasonable because they originate from bulk optical modes
that are highly Raman active. Since we have shown that these
shifts are caused by the strain and associated with structural
variations, the active Raman scattering peaks provide a convenient
way to obtain structural information and corresponding strain
condition inside Si/Ge core-shell NWs. As a result, they will be
able to answer an fundamental question in nanoscience, which is if
the huge strain can sustain in narrow core-shell NWs without
introducing an significant amount of dislocations.

We also observe that the Raman scattering signal intensity
strongly depends on the type of atoms involved in the vibrational
modes. As shown in Fig. \ref{fig5}, those Ge modes have much
stronger Raman signals than those Si modes. This can be attributed
to the larger size of Ge atoms, so that their motions can induce
much stronger changes of polarizability, leading to enhanced Raman
signals.

Meanwhile, the intensity of Raman scattering signals also depends
on the spatial location of those atoms involved in the vibrational
modes. The Raman scattering peak is usually enhanced if the
corresponding mode is within the core region while it is depressed
if the mode is within the shell region. For example, though SiGe-2
(Si core) NW has a lower proportion of Si atoms than GeSi-1
(Si shell) NW as listed in Table I, the relative Raman signal of the
highest-frequency Si mode in SiGe-2 NW is stronger than that in
GeSi-1 NW, as shown in Figs. \ref{fig5} (b) and (d). The similar
result can be observed in other NWs if we compare Figs. \ref{fig5}
(a) and (e).

Other than core-shell NWs, if comparing the Raman spectra of the
homogenous SiNW and GeNW with a similar diameter as shown in Figs.
\ref{fig5} (c) and (f), we find that the SiNW, different from the
GeNW which has only one dominant Raman peak at highest optical
frequency, displays bright Raman activities for many phonon modes
with lower frequencies. Since both bulk Si and Ge have only one
dominant peak in the first-order Raman scatter spectrum, those
newly active low-frequency modes in SiNWs imply that these
vibrational modes are more affected by the quantum confinement and
symmetry breaking in SiNWs. This has been observed by recent
experiments \cite{2000Wang,2005Adu}. Such a different variation of
vibrational modes may be helpful to explain why the thermal
conductivity of SiNWs differs from their bulk crystals more than
that of GeNWs \cite{2011Hu}.

\subsection{RBM}

After discussing the high-frequency optical vibrational modes in
Si/Ge core-shell NWs, we turn to another type of collective modes:
the RBM. Because of the unique geometry of quasi-one-dimensional
nanostructures, their RBMs are of great interests and importance.
For example, they are characteristic modes reflecting the geometry
of carbon nanotubes \cite{2004Fantini,2004SouzaFilho}. Recent
research \cite{2011Yang,2010Bourgeois} has also shown that the RBM
are Raman active in narrow semiconducting NWs and they are tightly
related to the diameter of NWs. Then an obvious question is
whether we can observe the similar RBM in core-shell NWs.

\begin{figure}
\centering
\includegraphics[width=1.0\columnwidth]{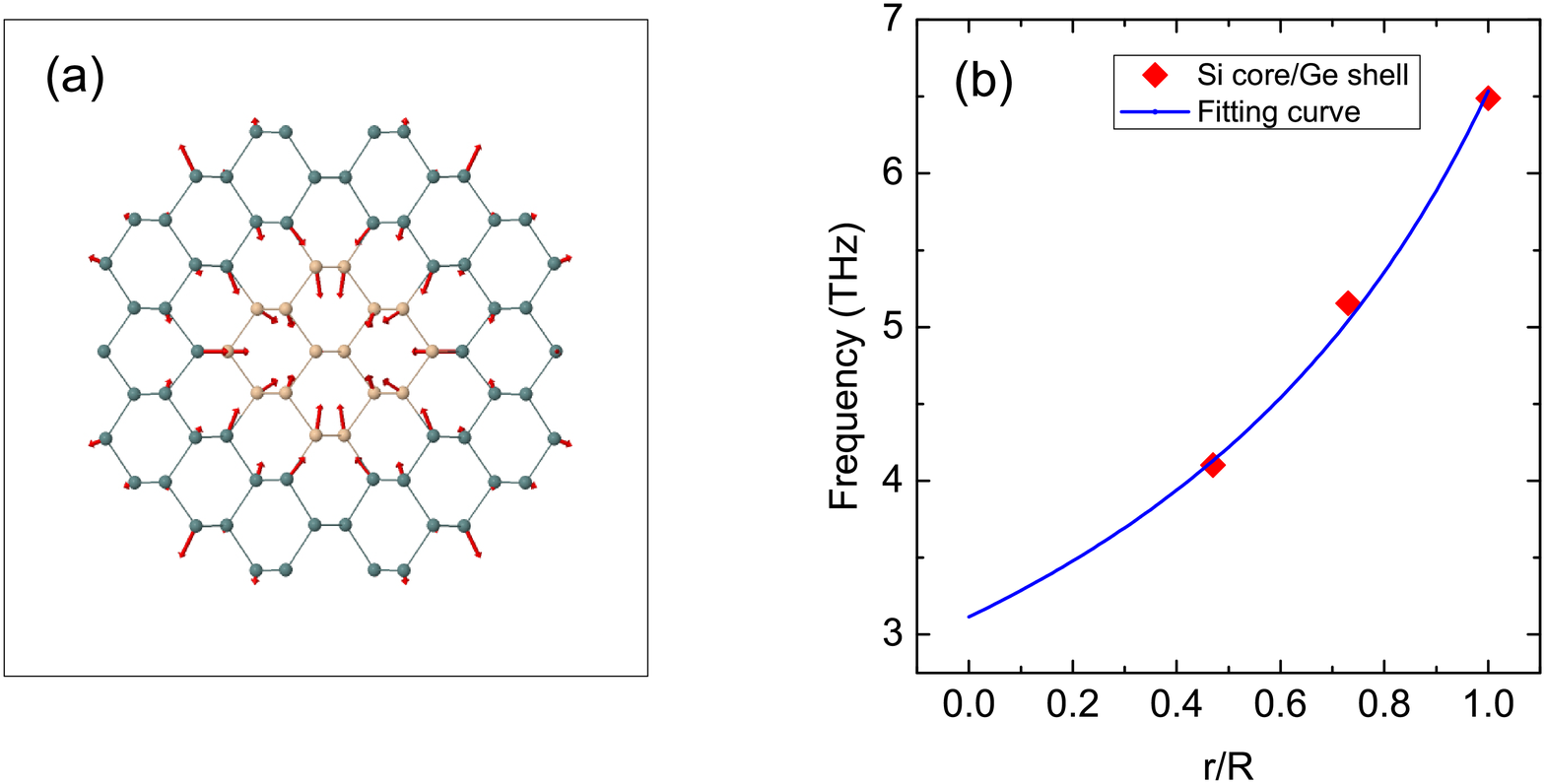}%
\caption{\label{fig6} (a) A typical RBM in Si/Ge core-shell NWs.
(b) The frequency of RBM according to the core/wire size ratio.
The fitting curve is based on Eq.~(\ref{equ8}).}
\end{figure}

Our first-principles result shows that the answer for core-shell
NWs is more complicated than usual homogenous nanostructures case.
For example, the RBM are identified in the Si-core/Ge-shell NWs
studied in this work as shown in Fig. \ref{fig6} (a). Moreover, we
find that this RBM are Raman active and can be detected by the
Raman scattering experiments, as marked in Fig. \ref{fig5} (a),
(b) and (c). On the other hand, for the other type of NWs,
Ge-core/Si-shell NWs, we cannot identify the RBM simply by sight.
In fact, even for GeNWs with a similar diameter as the studied
core-shell NWs, no RBM can be easily identified. The relative
softer bonding between Ge atoms and the relatively larger mass of
Ge atoms substantially reduce the frequency of the RBM. As a
result, the RBM will mix with other modes because of the
application of the sum rule in our simulations.

%This idea is reasonable because we do find the RBM in narrower Ge
%NWs (d $\sim$ 1.76 nm), with a frequency around 4.5 THz but cannot
%find it for wider GeNWs whose RBM shall have a lower frequency.

%In this sense, the RBM may not a good option to
%detect the structure of Si/Ge core-shell NWs.

The most important motivation to study the RBM is to see if they
have a strong relation with the structure of nanomaterials, such
as the diameter of NWs. Previous studies \cite{2011Yang} have
successfully applied an elastic media model to study the RBM in
SiNWs, encouraging us to apply a similar approach to study the RBM
in Si-core/Ge-shell NWs. By using the classic wave equation with a
cylindrical boundary condition and regarding core and shell
regions as two different homogenous elastic medias with different
sound velocities $c_{1}$ and $c_{2}$ respectively, the vibration
of RBM can be described as $\vec{u}(x,y,z) = \vec{u}(x,y) = u(r)
\hat{r}$ and satisfies the following equations:

\begin{subequations}
\label{equ5whole}
\begin{equation}
\frac{\partial^{2}u(r)}{\partial r^{2}} +
\frac{1}{r}\frac{\partial^{}u(r)}{\partial r} + (
\frac{\omega^{2}}{c_{1}^{2}}-\frac{1}{r^{2}})u(r) = 0  \;  for \;
0 \leq r \leq R_{1},
\end{equation}
\begin{equation}
\frac{\partial^{2}u(r)}{\partial r^{2}} +
\frac{1}{r}\frac{\partial^{}u(r)}{\partial r} + (
\frac{\omega^{2}}{c_{2}^{2}}-\frac{1}{r^{2}})u(r) = 0  \;  for \;
R_{1} \leq r \leq R_{2},
\end{equation}
\end{subequations}
where $\omega$ is the frequency of RBM. $R_{1}$ is the core radius
and $R_{2}$ is the shell radius, $u_{1}(r)$ and $u_{2}(r)$ are RBM
wave function in the core and shell region, respectively. We solve
Eqs.~(\ref{equ5whole}) using the free boundary condition at
$r=R_{2}$ and continuous and differentiable continuous condition
at $r=R_{1}$, we get the equation Eq.~(\ref{equ6}) which determine
$\omega$, the frequency of RBM:

\begin{widetext}
\begin{equation}\label{equ6}
{c_{1}}J_{1}(\frac{\omega
R_{1}}{c_{1}})[Y_{1}^{\prime}(\frac{\omega R_{1}}{c_{2}})
J_{1}^{\prime}(\frac{\omega R_{2}}{c_{2}})-
Y_{1}^{\prime}(\frac{\omega R_{2}}{c_{2}})
J_{1}^{\prime}(\frac{\omega R_{1}}{c_{2}})] +
{c_{2}}J_{1}^{\prime}(\frac{\omega
R_{1}}{c_{1}})[Y_{1}^{\prime}(\frac{\omega R_{2}}{c_{2}})
J_{1}(\frac{\omega R_{1}}{c_{2}})- Y_{1}(\frac{\omega
R_{1}}{c_{2}}) J_{1}^{\prime}(\frac{\omega R_{2}}{c_{2}})] = 0,
\end{equation}
\end{widetext}
where $J_{1}$ and $Y_{1}$ are first-order Bessel functions of the
first kind and second kind, respectively. By applying perturbation
theory and the approximation of Bessel's functions in
Eq.~(\ref{equbesselapprox}) when $x \gg \frac{3}{4}$ (it is
satisfied in our simulations),

\begin{subequations}
\label{equbesselapprox}
\begin{equation}
J_{1}(x) \approx \sqrt{\frac{2}{\pi x}} cos(x-\frac{3}{4}\pi),
\end{equation}
\begin{equation}
Y_{1}(x) \approx \sqrt{\frac{2}{\pi x}} sin(x-\frac{3}{4}\pi).
\end{equation}
\end{subequations}

Then a relation about the frequency of RBMs in core-shell NWs is
obtained:

\begin{equation}\label{equ7}
\omega
(\frac{R_{2}}{c_{2}}+\frac{R_{1}}{c_{1}}-\frac{R_{1}}{c_{2}})\approx
constant.
\end{equation}

Therefore, the frequency $\omega$ of the RBM is determined by the
ratio of core/wire size and that of the sound velocities of core
and shell. The constant in Eq.~(\ref{equ7}) are a series of values
of $x$ when $J_{1}^{\prime}(x)=0$, according to the free boundary
condition. If we assume the sound velocity does not change too
much, the above Eq.~(\ref{equ7}) can be approximated to

\begin{equation}\label{equ8}
\omega \propto 1/
[\frac{c_1}{c_2}+\frac{r}{R}(1-\frac{c_1}{c_2})].
\end{equation}

From the equation Eq.~(\ref{equ7}), we can easily reduce it to the
known RBM frequency of homogenous NWs \cite{2011Yang}:

\begin{equation}\label{equ9}
\omega \frac{R}{c}\approx constant.
\end{equation}

%We have applied the equation Eq.~(\ref{equ7}) to fit the data of
%RBM in the Si-core/Ge-shell NWs, as shown in Fig. \ref{fig6}(a).
%From the fitting curve, the phonon group velocity ratio in Si core
%and Ge shell is about 1.82, which is bigger than the ratio in
%homogenous NWs or bulk materials ($\sim$ 1.56). Thus, the
%Eq.~(\ref{equ7}) is useful to describe RBM frequency relation in
%core-shell NWs.

We have applied the equation Eq.~(\ref{equ8}) to fit the data of
RBMs in the studied Si-core/Ge-shell NWs, as shown in Fig.
\ref{fig6}(b). From the fitting curve, the ratio of phonon group
velocities in Si core and Ge shell is about 1.82, which is larger
than the ratio in homogenous NWs or bulk materials ($\sim$ 1.56).
This slight enlargement of the phonon group velocity ratio in
core-shell NWs is a result of intrinsic strain effect and it may
be of interest for thermoelectric applications of core-shell NWs.

\section{\label{sec4}CONCLUSIONS}

In summary, we have studied lattice vibrational modes of Si/Ge
core-shell NWs by the first-principles approach. The frequency of
high-energy Si-Si and Ge-Ge optical modes exhibits significant
shifting when the core-shell size varies. Our analysis shows
that these shifts are related to the variation of the strain
conditions inside core-shell NWs. At the same time, the RBM are
identified in the Si-core/Ge-shell NWs, whose frequency strongly
depends on the geometry of NWs, and our elastic media model explains
well the frequency of the RBM. Moreover, we have performed the
calculation to obtain the first-order Raman scattering spectra of
studied NWs, in which the shift of high-frequency optical modes and
the RBM can be identified, providing a convenient way to detect the
structural information of core-shell NWs. Our studied vibrational
modes and their frequencies could be useful for thermoelectric
applications as well.

\begin{acknowledgments}
Support from the International Center for Advanced Renewable
Energy and Sustainability (I-CARES) is gratefully acknowledged. We
acknowledge computational resource support by the Lonestar of
Teragrid at the Texas Advanced Computing Center (TACC) and the
National Energy Research Scientific Computing Center (NERSC)
funded by the U.S. Department of Energy.
\end{acknowledgments}


\end{document}